\documentclass[prb, twocolumn, secnumarabic, amsfonts, amssymb, amsmath, aps, byrevtex, showpacs]{revtex4-1}
\usepackage{booktabs}  
\usepackage{amsmath}
\usepackage{mathrsfs}
\usepackage{xcolor}
\usepackage{graphicx}    
\usepackage{ctable}
\usepackage{longtable}

\begin{document}

\title{Fixed-Node Diffusion Monte Carlo of Lithium Systems}%

\author{K. M. Rasch}%
\email{kmrasch@ncsu.edu}
\affiliation{Physics Dept., North Carolina State University}
\author{L. Mitas}
\email{lmitas@ncsu.edu}
\affiliation{Physics Dept., North Carolina State University}
\begin{abstract}
We study lithium systems over a range of number of atoms, e.g., atomic anion, dimer, metallic cluster, and body-centered cubic crystal by the diffusion Monte Carlo method.
The calculations include both core and valence electrons in order to avoid any possible impact by pseudopotentials.
The focus of the study is the fixed-node errors, and for that purpose we test several orbital sets in order to provide the most accurate nodal hypersurfaces.
We compare our results to other high accuracy calculations wherever available and to experimental results so as to quantify the the fixed-node errors.
The results for these Li systems show that fixed-node quantum Monte Carlo achieves remarkably
high accuracy total energies and recovers 97-99 \% of the correlation energy. 
\end{abstract}

\maketitle
\tableofcontents

\section{Introduction}
Quantum Monte Carlo (QMC) methods have been applied to a great variety of electronic structure problems over the past three decades.
These calculations have provided a number of high accuracy results for properties such as cohesion and binding energies, excitations, reaction barrier heights, defect formation energies, and other quantities; they are typically in excellent agreement with available experiments~\cite{Foulkes:2001uq, Kolorenc:2011fk}.
In addition, the calculations have shed new light on correlation effects in various systems, and therefore have become valuable as benchmarks for other methods and comparisons.
The most important strength of this approach is that the many-body Hamiltonian is employed directly, and thus the electron-electron interaction and particle correlations are treated explicitly in a many-body manner.
Additionally, QMC methods can be applied to large systems so that properties of solids can be calculated by using supercells and extrapolations to the thermodynamical limit.

Diffusion Monte Carlo (DMC) projects out the ground state of a system by applying the projection operator $\exp(-\tau H), $ where $H$ is the Hamiltonian, to the trial wave function $\Psi_T$.
In the large imaginary time limit $\tau \to \infty$, the ground state of a given symmetry is obtained.
One of the key fundamental limitations to achieving exact results is the so-called fixed-node approximation which enables one to avoid the well-known fermion sign problem~\cite{Reynolds:1982fk, Foulkes:2001uq}.
The fixed-node approximation is very difficult to improve upon since the corresponding energies are typically very small, e.g., a few percent of the correlation energy where the correlation energy is itself a small fraction of the total energy.
Therefore systematic  improvements of the nodes through minimization of the total energy or variance of the energy~\cite{Reboredo:2009uq, Umrigar:2005fk} for a given trial wave function is laborious and often very costly.
Alongside the obviously desirable achievements needed in improving the algorithmic efficiency, in increasing the speed and quality of wave function optimizations, and in exploring new functional forms for orbitals and wave functions, a significant amount of work remains to be done in simply understanding the root origin of the nodal errors by systematically quantifying the dependence of energy biases on the nodal defects.
Recently, we have analyzed the impact of the electron density on the nodal bias in a set of free atom/ionic systems, and we found the fixed-node errors being proportional to the density in this particular class of systems for both spin-polarized and spin-unpolarized cases~\cite{Rasch:2012uq, Kulahlioglu:uq}.
More insights have been gained on the role of the basis sets~\cite{Umrigar:2005fk} in the error although the findings were not always easy to utilize in different systems~\cite{Bertini:2004fk}.

As a testbed to expand this study for more complicated cases, 
a series of lithium systems is attractive for several reasons. 
First, inclusion of the core electrons in Li calculations is computationally feasible. That enables us to avoid any additional, more complicated analysis that is necessary whenever pseudopotentials or effective core potentials are being employed.
This ensures that any missing amount of the binding energy, cohesive energy, or correlation energy is caused solely by the fixed-node approximation.
Second, the exact wave function for a free Li atom has a relatively simple nodal surface which is already well approximated at the Hartree-Fock level.
In addition, small Li systems have been studied with the FN-DMC method before. 
In this work we significantly expand upon previous studies with calculations of Li$_4$ cluster and Li solid in its equilibrium body-centered cubic structure.
By comparing our fixed-node diffusion Monte Carlo (FN-DMC) results with other accurate calculations and experimental results corrected for zero-point motion (DMC is carried out in the Born-Oppenheimer approximation), we can assess the magnitude of the fixed-node errors with high accuracy.
These systems represent a variety of environments for the bonding and include directional bonds, multi-center bonding, and delocalized metallic bonding.
It is therefore an interesting question to understand how the fixed-node bias changes once Li enters bonding in the setting of molecular bonds or periodic boundary conditions. 
Based on these results we establish a systematic picture of the nodal errors in Li systems and corresponding accuracies for energy differences.

\section{Methods}
In DMC, we solve for the ground state solution of Schr\"odinger's equation
\begin{equation}
\Psi_0 = \lim_{\tau \rightarrow \infty} \exp \{ - \tau \mathscr{H}\} \Psi_\text{T}
\end{equation}
where $\mathscr{H}$ is the Born-Oppenheimer Hamiltonian.
The antisymmetric nature of fermion systems poses a challenge to the naive application of the DMC algorithm and leads to the fermion sign problem ~\cite{Anderson:1976uq, Reynolds:1982fk, Zhang:1991fk}.
This is because for a given boundary value problem, the eigenstate with the lowest eigenvalue will be a symmetric state.
In light of this well-known difficulty, perhaps the simplest and most straightforward way to circumvent the sign problem is the fixed-node approximation.

Under the fixed-node approximation, we impose a boundary condition at the nodes of the trial wave function and maintain them for the duration of the simulation. 
The nodes form a hypersurface defined implicitly by
\begin{equation}
\Omega = \{ \mathbf{R} ; \Psi_\text{T}(\mathbf{R}) = 0 \}
.
\end{equation}
The assumed nodal hypersurface creates boundaries that constrain the solution in each nodal cell and preserve the overall fermionic anti-symmetry of the total wave function, preventing thus any appearance of ``signs.''
This allows one to ignore the sign of the wave function inside the nodal cell and to carry out the DMC algorithm within each nodal cell
\begin{equation}
\Psi_\text{T} \Psi_0 \geq 0
.
\end{equation}
Unfortunately, doing this exactly is a tall order as it requires that for an $N$ electron system one must have a description of the exact $\left(3N-1\right)$-dimensional hypersurface $\Omega$.
Solving for such a hypersurface directly is beyond our means, and instead we proceed by using nodal surfaces from approximate wave functions.
Because we use a nodal hypersurface that is not exact, the solution will have a higher energy than the exact ground state, i.e., the total energy computed via FN-DMC is a variational upper bound to the exact energy~\cite{Moskowitz:1982uq}. 
Further details on the FN-DMC method can be found elsewhere~\cite{Foulkes:2001uq}.

The trial functions used in this study are of the Slater-Jastrow type 
\begin{equation}
\Psi_\text{T}({\bf R})= \sum_kc_k {\rm det}_k^{\uparrow}[\varphi_i]
{\rm det}_k^{\downarrow}[\varphi_j] \exp(U)
\end{equation}
where the one-particle orbitals are obtained from Hartree-Fock (HF) or Density Functional Theory (DFT). 
More details on the trial functions and their optimizations, Jastrow correlation factors and DMC calculations for periodic boundary conditions can be found in the recent review~\cite{Kolorenc:2011fk}.

\section{Results}
\subsection{Lithium atom \& electron affinity}
In order to calculate the total energy of the lithium atom, we use a restricted Hartree-Fock, 
single-determinant wave function. 
Since at the HF level the spin-up and spin-down subspaces are independent,
the minority spin channel contains only one electron, and the wave function's node exists in only in the spin majority subspace.
We can visualize this subspace of the nodal hypersurface by considering the wave function when 1 and 2 label the electrons in the same spin channel.
Then it follows from the form of the HF determinant that the node is given by the condition $r_1= r_2$.
The electron 1 therefore ``sees'' the node as a sphere which passes through the position of electron 2 and is centered around the nucleus.
The wave function will be equal to zero if electron 1 occupies any point on the spherical nodal surface.

The exact HF nodal hypersurface in the full 6D space is a 5D hyperboloid given by the implicit equation $x_1^2 + y_1^2 + z_1^2 = x_2^2 + y_2^2 + z_2^2$.
As pointed out by Stillinger \emph{et al.}~\cite{White:1971fk}, this is not strictly exact, as the correlation with the electron in the spin-down channel will cause deformations away from a perfect sphere.
For example, the excitation $2s^12p^2$ will have a contribution to the exact ground state and would in principle lead to a departure from the single particle node (i.e., the sphere will slightly deform to ellipsoid or perhaps a more complicated surface that would depend on the position of the minority spin electron).
It is therefore quite remarkable that the HF nodal surface seems to be so accurate: the total energy with the HF nodes, shown in Table~\ref{table:lithiumAtom}, is accurate to $\approx0.05(1)$~mHa and the fixed-node bias is less than 0.1~\% of the correlation energy~\cite{Bressanini:2002fk}.
This points out that the correlation is basically completely captured by the Jastrow-like effect and it affects the 5D hyperboloid only marginally. (This contrasts with the Be atoms where the nodal surface is strongly affected by correlations~\cite{Rasch:2012uq}.)
\begin{table}
\caption{Comparison of theoretical results for the total energy of a lithium atom where FN-DMC energy has been obtained with the HF nodes.}
\begin{center}
\begin{tabular}{ c  l }
\hline  \hline
source & total energy (Ha)\\
 \hline \hline
FN-DMC present work                                                               & -7.47801(1)\\
FN-DMC Bressanini \emph{et al.}~\cite{Bressanini:2002fk}	& -7.47803(5)\\
exact~\cite{Yan:1998fk}                                              & -7.47806032\\
\hline
\end{tabular}
\end{center}
\label{table:lithiumAtom}
\end{table}%
Our own calculated fixed-node error in the single atom energy is $\approx 0.05(1)$ mHa, much smaller than, say, the chemical accuracy ($\approx1.6$ mHa).
This also suggests that any fixed-node errors in the aggregate species from Li atoms will be essentially identical to the error in the binding or cohesive energy.

Using the value for the ground state total energy of the Li$^{-}$ ion from reference~\cite{Rasch:2012uq}, which has a fixed-node error comparable to the neutral atom, we can compute the electron affinity of a Li atom.
The electron affinity is given by
\begin{equation}
\text{EA}(\text{Li}) = E_0(\text{Li}) - E_0(\text{Li}^-)
.
\end{equation}
Lithium has a positive electron affinity, meaning the anion is more stable than the neutral atom.
The HF limit of the total energy of Li$^-$ has been computed~\cite{Agren:1989uq} to be $-7.4282320$ Ha so that in the HF approximation the additional electron is not bound. As it is well-known, the correlation effects
 are crucial for describing the electron affinity with accuracy comparable to experiment ~\cite{King:1996kx}.

Using a wave function composed of two configuration state functions, i.e., HF plus $2s^2\to2p^2$ excitation (a symmetry adapted linear combination of determinants), given the data from our recent calculations~\cite{Rasch:2012uq} for the 4 electron ion yields an FN-DMC electron affinity with an excellent accuracy compared to experimental measurement~\cite{Haeffler:1996vn}.
A summary of the best theoretical and experimental data are compared with this work in Table~\ref{table:ElectronAffinity}.
\begin{table}
\caption{Comparison of the latest calculation and measurement with FN-DMC results for the electron affinity for Li (in Ha).}
\begin{center}
\begin{tabular}{ l l l }
\hline \hline
Author&Method&EA (Ha)\\
\hline \hline
Fischer~\cite{Fischer:1993ys}&extrap. MCHF& 0.022698\\
\emph{present work} &FN-DMC HF single det.  & 0.0201(1)\\
\emph{present work} &FN-DMC 2 configs.  & 0.02279(5) \\
Haeffler~\cite{Haeffler:1996vn} &\emph{expt.}&0.0227129(8) \\
\hline
\end{tabular}
\end{center}
\label{table:ElectronAffinity}
\end{table}%
In terms of the fixed-node error, while the two species in the calculation share geometric details (central potential in free space), the nodal hypersurface changes when the 4th electron is added and the anion
shows similar nodal shape as the isoelectronic Be atom. 
The poor quality of the result for the electron affinity using only a single determinant trial function for Li$^-$ stands to illustrate that the extreme accuracy of the RHF nodal hypersurface for 3 electrons is 
not typical and rather results from a fortuitous coincidence.

\subsection{Li$_2$}
The Li$_2$ dimer is a more complicated system, with increased number of electrons while also changing the overall real-space geometry from one central potential with spherical symmetry to a cylindrical symmetry.
The nodes of the lithium dimer have been a studied number of times~\cite{Filippi:1996fk, Bressanini:2005bs}.
The best result in the literature (in reference~\cite{Bressanini:2005bs}) exceeds the quality of our single configuration result, $E_\text{FN-DMC} = -14.9932(7)$, which has almost $2$~mHa fixed-node error.
The fixed-node error of the best wave function in the literature is~$\approx~0.2~-~0.3~$~mHa, with a total energy of $-14.9952(1)$.
Using the value for the HF energy reported by Filippi and Umrigar~\cite{Filippi:1996fk}, this recovers~$\approx 99.8\%$ of the correlation energy.
Bressanini \emph{et al.}~\cite{Bressanini:2005bs} have pointed out that a five configuration wave function basically gets the nodal surface close to exact, and one can get accurate energies at the fixed-node approximation level with errors of the order of the order of 0.1 \% as well.
Interestingly, only the three lowest excitations were really involved, $2\sigma^2 \to 1\pi^2$, $2\sigma^2_g \to 2\sigma^2_u$, $2\sigma^2_g \to 3\sigma^2_g$.
Since this system was studied exhaustively, we did not repeat the muti-reference calculation, and instead we quote the results of Bressanini \emph{et al.}~\cite{Bressanini:2005bs} hereafter.

\subsection{Li$_4$}
\begin{figure}
	\centering
	\includegraphics[width=\columnwidth]{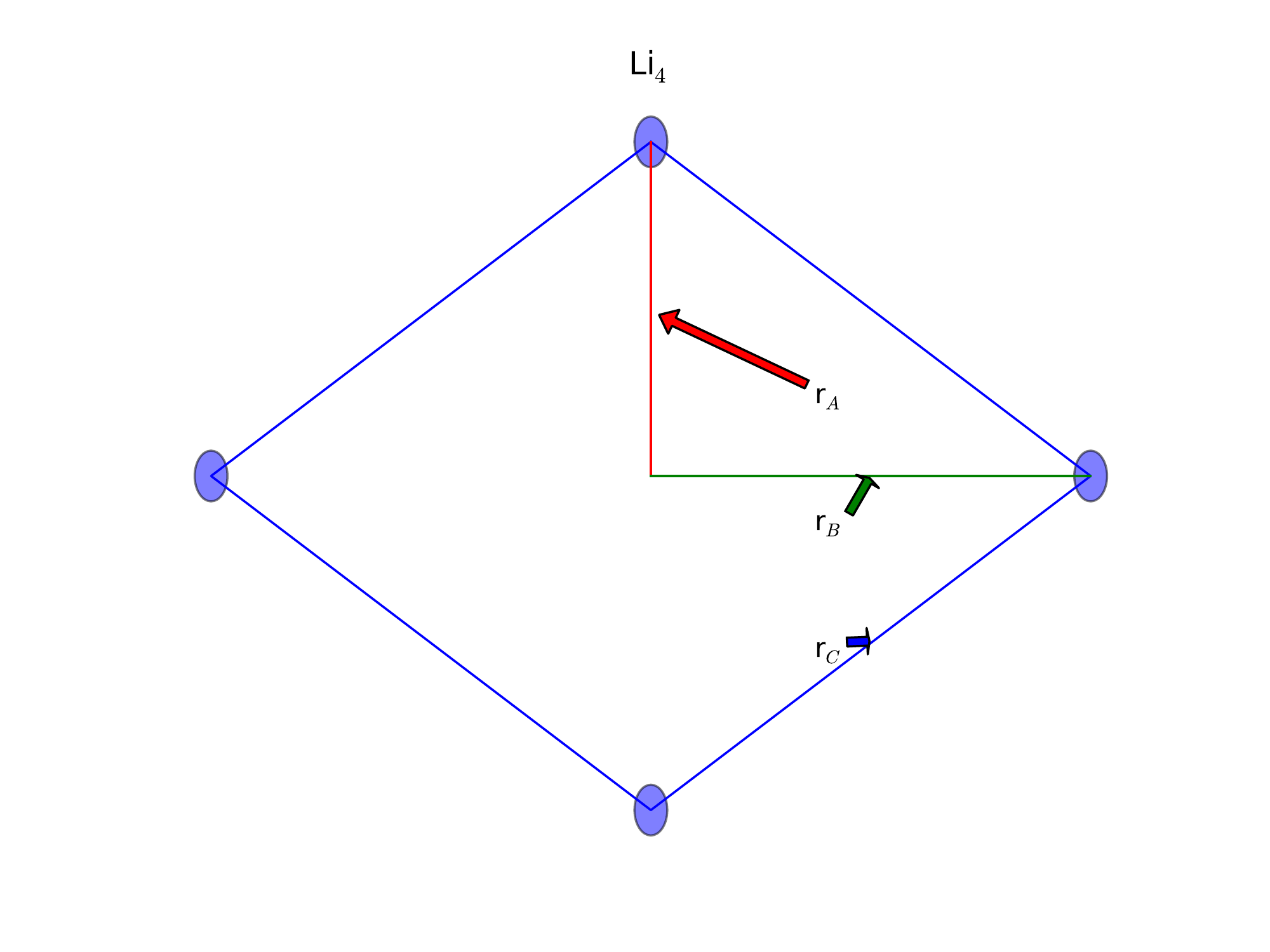}
	\caption{Schematic depiction of the $D_{2h}$ Li$_4$ parameters}
	\label{fig:li4geom}
	\end{figure}

Li$_4$ and its properties have been studied by several methods including the basis set correlated approaches~\cite{Ray:1985zr, Boustani:1987ly, Jones:1997ve, Rousseau:1997qf, Wheeler:2004bh, Nissenbaum:2009dq}.
The most stable configuration of four Li atoms is a molecule with $D_{2h}$ symmetry, a planar rhombus geometry, and singlet electronic ground state~\cite{Ray:1985zr, Boustani:1987ly, Jones:1997ve, Rousseau:1997qf, Wheeler:2004bh}.
The geometry of the $D_{2h}$~Li$_4$ is depicted schematically in Figure~\ref{fig:li4geom}.
This can be understood as the result of a Jahn-Teller distortion of the more symmetric geometry of a square~\cite{Jones:1997ve}.
Li$_4$ exhibits a ``three-center'' bonding pattern where two electrons are shared inside each of the two triangles formed by bisecting Figure~\ref{fig:li4geom} along the ``r$_A$'' line~\cite{Rousseau:2000fv}.

In Table~\ref{table:li4nodalcomparison}, we compare the nodes of the SCF wave functions (unoptimized) for different levels of CI in order to illustrate the behavior of such expansions and to select the best starting place for our QMC trial wave function.
For each level of CI, we used complete expansions but limited the number of virtual orbitals in the active space.
\begin{table}
\caption{Fixed-Node DMC total energy for trial wave functions from different levels of theory testing unoptimized nodal surfaces for use as DMC trial wave-functions.}
\begin{center}
\begin{tabular}{ l c l }
\hline \hline
theory & size of virtual space & $E_\text{tot}^\text{FN-DMC}$ (Ha) \\
\hline \hline
RHF		&	0	& -30.0177(5)	\\
\hline
CI-SD	&	9	& -30.0184(5)	\\
CI-SDTQ	&	9	& -30.0174(4)	\\
CI-SD	&	15	& -30.0228(4)	\\
CI-SDTQ  &	15	& -30.0179(4)	\\
CI-SD	&	19	& -30.0162(4)	\\
CI-SDTQ	&	19	& -30.0179(6)	\\
\hline
\end{tabular}
\end{center}
\label{table:li4nodalcomparison}
\end{table}%
It is clear that the nodes do not improve systematically for larger active space and higher level of theory as the CI total energies do. 

Since FN-DMC errors associated with the basis set are not very systematic, we also tested the nodal surfaces of several basis sets to minimize these errors.
Although not fully complete, the results seem to support the conclusion of Bressanini~\emph{et al.} that for Li systems saturating the $s$ channel is more important than adding additional high angular momentum basis functions.
These results are listed in Table~\ref{table:li4BasisSet}.
\begin{table}
\caption{FN-DMC results for different basis sets with trial wave functions from CI-SD calculations using 15 virtual orbitals and then optimized in with respect to VMC total energy.}
\begin{center}
\begin{tabular}{ l  l }
\hline \hline
basis set 					& total energy (Ha)\\
\hline \hline
Roos Aug. DZ ANO	(4s3p2d)		&-30.02127(5)\\
Roos Aug. TZ ANO 	(4s4p3d2f)	&-30.02119(6)\\
aug-cc-pCVTZ 		(7s6p4d2f)	&-30.02263(6)\\
\hline
\end{tabular}
\end{center}
\label{table:li4BasisSet}
\end{table}%
After some initial testing of basis and multi-determinant expansions, we employed wave functions constructed from the aug-cc-pCVTZ basis and included the 15 lowest lying virtual orbitals into the CI-SD calculation.
We re-optimized the weights of the resulting 93 configuration state functions in the CI expansion with VMC total energy minimization using a Levenberg-Marquardt algorithm.
The geometry parameters have been computed a number of times in the literature, as reported in Table~\ref{table:li4Geom} organized by the value for the Li-Li distance labeled $r_C$ in Figure~\ref{fig:li4geom}.
\begin{table*}
\caption{Summary of the optimized geometry parameters of $D_{2h}$ Li$_4$ tested in this work, and the FN-DMC total energy for each.  The trial wave function is an VMC energy optimized CI-SD expansion with 93 CSFs.}
\begin{center}
\begin{tabular}{ l l l l l  l }
\hline \hline
author 									& method 			&r$_A$ (\AA)&r$_B$ (\AA)&r$_C$ (\AA)& $E_\text{tot}^\text{FN-DMC}$ (Ha)\\
\hline \hline
Ray~\cite{Ray:1985zr}				 		&DFT			& 1.298 & 2.759 & 3.050 & -30.02132(2)   \\
Rousseau and Marx~\cite{Rousseau:1997qf} 		&QCISD / CCSD(T)	& 1.323 & 2.700 & 3.007 & -30.02155(2)   \\
Verdicchio \emph{et al.}~\cite{Verdicchio:2012ij} 	&CCSD(T)		& 1.323 & 2.680 & 2.989 & -30.02158(2)   \\
Wheeler \emph{et al.}~\cite{Wheeler:2004bh} 		&CCSD(T)		& 1.316 & 2.681 & 2.987 & -30.02154(2)   \\
\hline
\end{tabular}
\end{center}
\label{table:li4Geom}
\end{table*}%
The results indicate that there is a short Li-Li bond $\approx2.64$~\AA\ (2r$_A$ in Figure~\ref{fig:li4geom}) and a longer Li-Li bond $\approx2.99-3.0$~\AA\ (r$_C$ in Figure~\ref{fig:li4geom}).

The experimental and theoretical binding energies of Li$_4$ are given in Table~\ref{table:li4}.
Because the presence of the Jastrow factor will influence the optimization of the multi-determinant expansion, it is not clear \emph{a priori} what form of the Jastrow factor is optimal. In particular,
since the bonds are not very strong and the bond lengths are somewhat larger than in typical
single-bonded situations we tested the range of the Jastrow cutoff distance parameter.
We optimized the Jastrow coefficients and determinant weights for two different electron-ion Jastrow distances.
The wave function with the so-called ``short-range'' Jastrow effects had its electron-ion and 
electron-electron-ion terms extend to 2.45 bohr from each atom, i.e., to just less than half 
the smallest Li-Li distance.
For the ``long range'' Jastrow, the electron-ion terms were allowed to extend to 7.5 bohr.
The qualitative difference between these two Jastrows is that terms from different atoms in the short-range Jastrow do not overlap in the region occupied by the three-center bonds.
This difference in description of the wave function translates into a difference in computational effort.
In the ``short-range'' case, each electron will have non-zero three-body Jastrow terms associated with only 1 atomic center at any given time; whereas in the ``long-range'' case, electrons have non-zero contributions to the Jastrow coming from each of the Li atoms surrounding the three-center bonding region.
The resulting effect in the total energy of the wave function was only $\approx 0.00013$ per atom.

\begin{figure}
	\centering
	\includegraphics[width=\columnwidth]{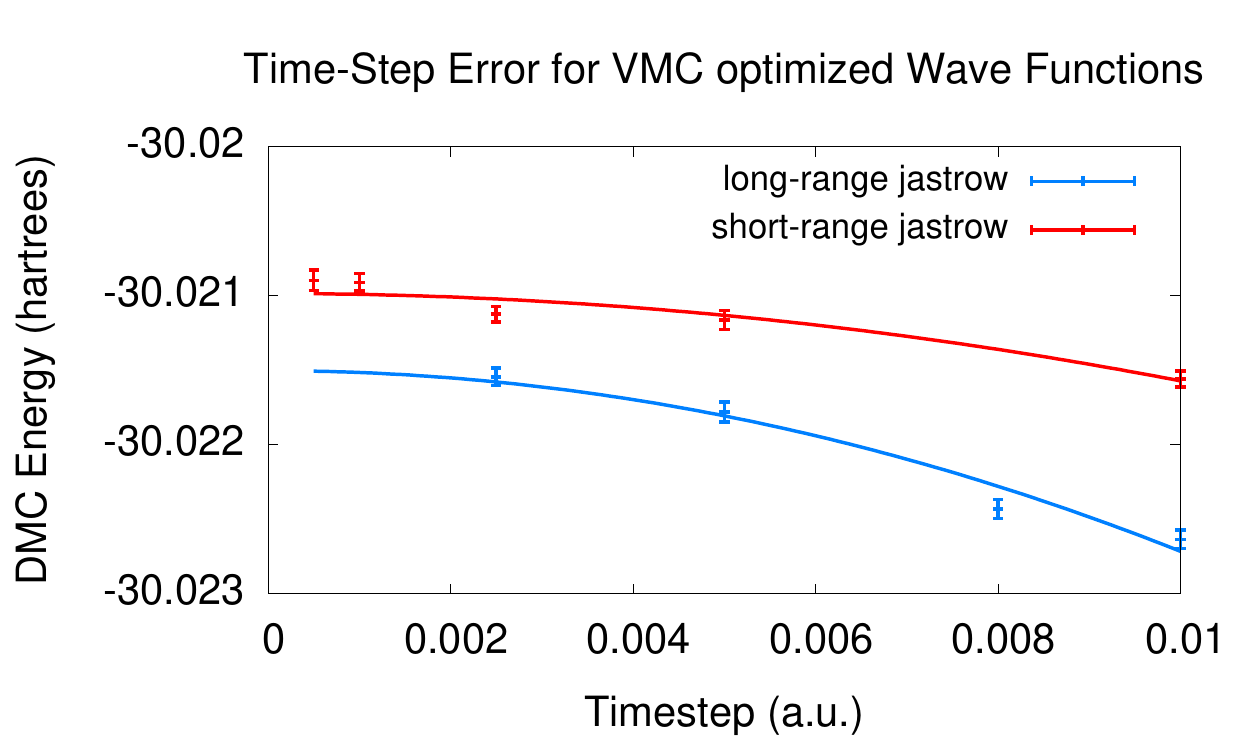}
	\caption{The DMC energy extrapolated to $\tau=0$ for a Li$_4$ molecule.  Both wave functions originate from the same Configuration Interaction CI-SD calculation, but have had the determinant weights and Jastrow coefficients re-optimized with different assumed cut-off distances in the Jastrow, as described in the text.}
	\label{fig:timestepError}
	\end{figure}

We carried out a time-step extrapolation for both wave functions, shown in Figure~\ref{fig:timestepError}, to ensure that the time-step error is $< \approx 0.05$~mHa in the total energy, or an order of magnitude smaller than the statistical error bars in the binding energy.
After correcting for the zero point motion, which was about 3.12 mHartree per atom~\cite{Wheeler:2004bh}, we find a binding energy of 0.723(3) $e$V.
\begin{table}
\caption{Binding energies of Li$_4$, uncorrected for zero-point motion, are given in units of $e$V per atom.}
\begin{center}
\begin{tabular}{l l c}
\hline \hline
author					& method 						& binding\\
\hline \hline
Alikhani and Shaik~\cite{Alikhani:2006hc}					& DFT		& 0.61\\
Bonacic-Koutecky \emph{et al.}~\cite{Bonacic-Koutecky:1993nx}	& MRD-CI 	& 0.63\\
Owen~\cite{Owen:1990tg}								& DMC		& 0.67(2)\\
Nissenbaum \emph{et al.}~\cite{Nissenbaum:2009dq}			& DMC		& 0.733(4)\\
Rao \emph{et al.}~\cite{Rao:1985kl}							& CI-SD		& 0.7375\\
Wheeler \emph{et al.}~\cite{Wheeler:2004bh}					& CCSD(T)	& 0.7445\\
\hline
present work											& DMC		& 0.744(3)\\
\hline
Wu~\cite{Wu:1983cr}									& expt.		& 0.84(5)\\
Brechignac \emph{et al.}~\cite{Brechignac:1994oq}				& expt.		& 0.63(4)\\
\hline
\end{tabular}
\label{table:li4}
\end{center}
\end{table}%
Note the reasonably good agreement between the results although the basis set correlation methods did not include any correlation of the core ($1s$) states.
This points out that the core states are already quite deep and do not affect the nodal surfaces significantly.
One of the reasons is that any excitation which would correlate the $1s$ level would involve states which would lie very high in energy since such excitations would require strongly localized type of orbitals. 
The fact that the nodal surfaces are minimally affected by the $1s$ sub-shell is also suggested by the highly accurate calculations presented above for Li  dimer.

The table includes also experimental data from the two available sources, Wu~\cite{Wu:1983cr} and Brechignac \emph{et al.}~\cite{Brechignac:1994oq}, 0.84(5) and 0.63(4)~$e$V respectively.
These data show significant differences and seem inconsistent with each other.
Considering a reasonable agreement between the four independent calculations in Table~\ref{table:li4}, we essentially claim that the experimental data is very inaccurate and our present calculation, being produced by an upper bound method and having the lowest total energy of the theoretical calculations, is the most accurate estimation of this total energy to date.

\subsection{Bulk Lithium in a BCC crystal}
Because of its position as the bulk crystal with only one valence electron per atom, lithium in the body-centered-cubic (BCC) crystal (Pearson symbol cI2) has been studied a few times by QMC methods in the literature~\cite{Sugiyama:1989uq, Eckstein:1995vn, Yao:1996kx, Filippi:1999ys}.
Surprisingly though, none of these calculations have used FN-DMC with the core electrons included.

Further interest in studying lithium crystals with QMC methods was stimulated by recent experimental and theoretical developments. For example, interesting phenomena for high pressures, including superconductivity, have been reported by experimental studies
~\cite{Hanfland:2000ly,Deemyad:2003zr}.
An intriguing hypothetical suggestion has been raised by Neaton and Ashcroft that lithium solid may undergo a Peierls transition into a so-called ``alkali electride'' at high pressures, an exotic phase where paired Li atoms are stabilized by pockets of highly localized electrons~\cite{Neaton:1999ve, Pickard:2009qf}.

\begin{figure}
	\centering
	\includegraphics[width=\columnwidth]{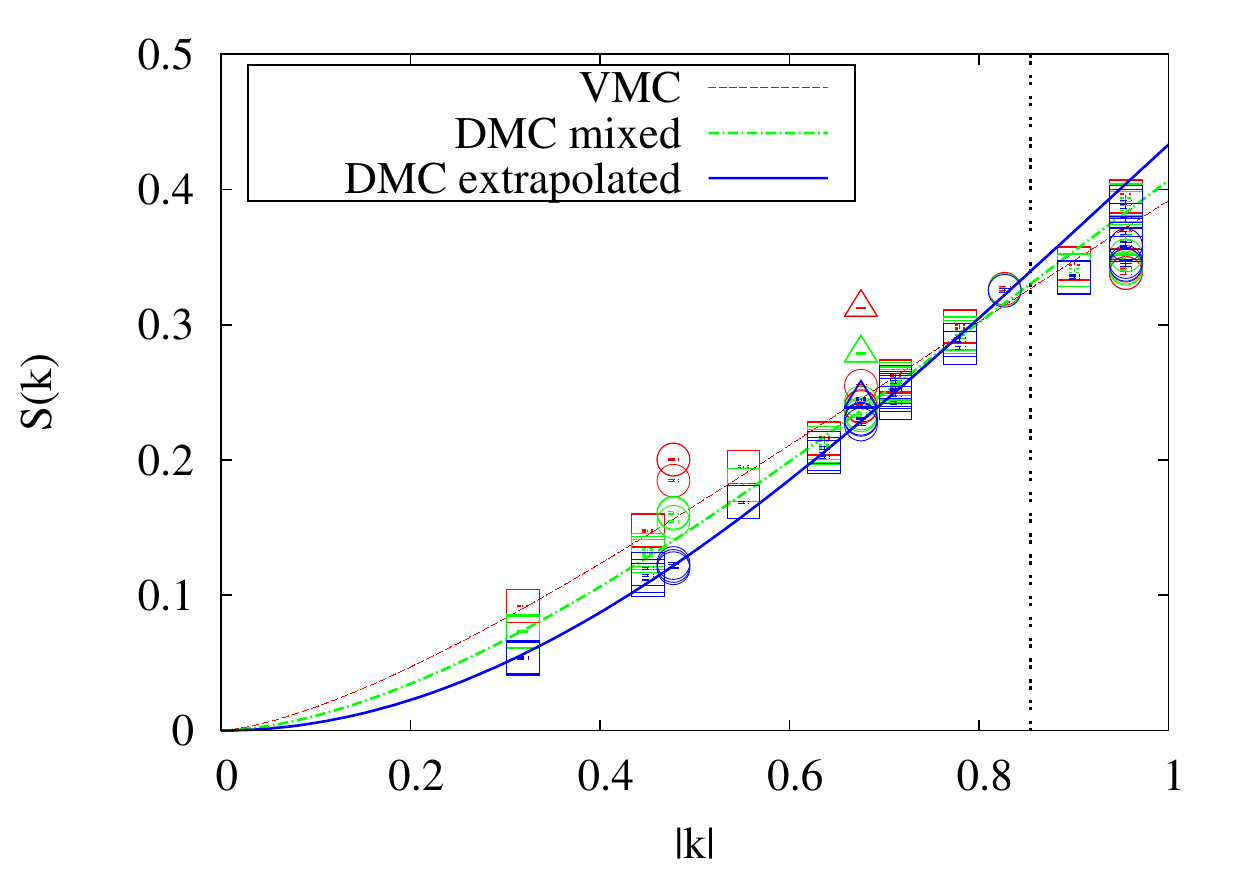}
	\caption{QMC results for $S(k)$ for several sizes of simulation cell.  Square symbols denote results from the 54-atom simulation; circle symbols, 16-atom simulation; and triangle symbols, 8-atom simulation.  The curves shown are fit to the 54-atom data with the function $S(k) = \exp\left\{ -a k^b\right\}$.    For small $k$, RPA predicts $b=2$.  For the fits shown, the $b$ parameter for VMC are 1.45; DMC with the mixed estimator, 1.67; and the DMC extrapolated estimator, 1.99.
	}.
	\label{fig:structureFactor}
	\end{figure}

\begin{figure}[h!]
	\includegraphics[width=\columnwidth, 
				   viewport= 0 0 0 0,
	                               trim= 0 0 0 0,
	                               clip=true]{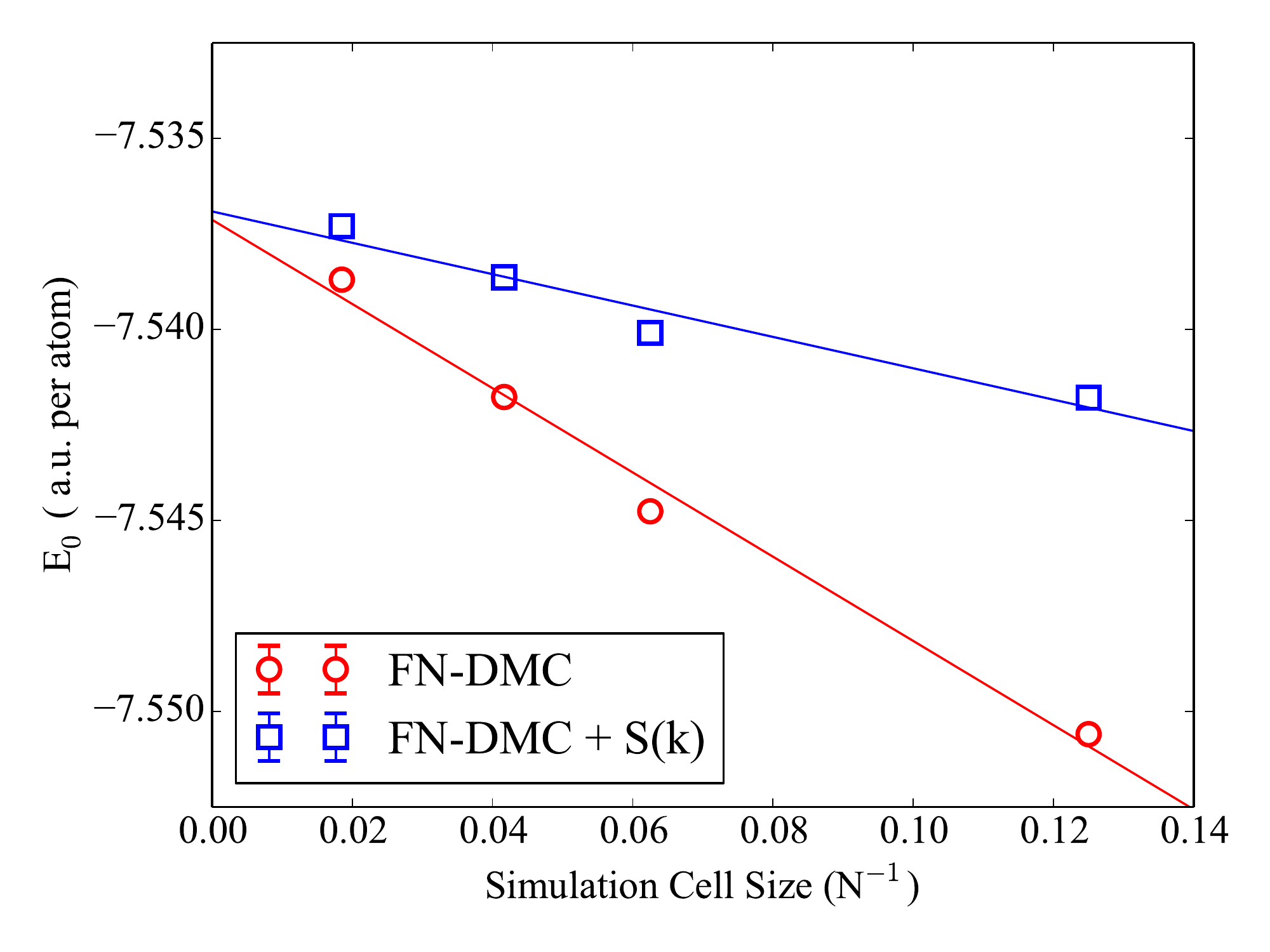}
	\caption{ The total energy per atom for twist averaged FN-DMC with and without the finite size error (FSE) corrections~\cite{Chiesa:2006uq} plotted against the inverse of the number of atoms in the cell.  The data are fit with an extrapolation to infinite bulk size.  The infinite bulk total energy per atom of -7.5371 Ha for the FN-DMC, and -7.5369 for FN-DMC with FSE corrections.  These values are within error bars of each other.  The statistical error bars on the data are smaller than the size of the plot symbol and so are not visible.  
	\label{fig:liCrystalSizeExtrap}
	}
	\end{figure}

We treat lithium solid in the cI2 crystal by FN-DMC at the experimental lattice constant~\cite{Filippi:1999ys, Vaidya:1971uq} $a_0 = 3.482$~\AA.
Our FN-DMC calculations use a single determinant Slater-Jastrow wave function with orbitals taken from DFT calculations.
Since the nodal surface would be rather difficult to improve upon within the QMC calculation of crystalline system, we begin by comparing several DFT functionals to find the best nodal surface; and we found the results to lie within $2-2.5$~mHa per atom of the highest quality nodal surface (PBE-PZ functional).
This strategy has been motivated by our work on DFT generated orbital sets to find the most optimal nodal surface~\cite{Kolorenc:2008uq, Kolorenc:2010fk}.
A select subset of the functionals tested are reported in Table~\ref{table:bulkDFTNodalQuality}.
\begin{table}
\caption{Results of FN-DMC calculations for the $\Gamma$-point wave function of an 8 atom supercell comparing the nodal quality of orbital sets generated by DFT functionals.}
\begin{center}
\begin{tabular}{ l  l  l }
\hline \hline
          Exchange&Correlation&$E_\text{tot}$ (Hartrees)\\
\hline \hline
          H-F                   &                           & -60.135(1)\\
          PW91-GGA     & PW91-GGA     & -60.136(1)\\
          PBE-GGA        & PBE-GGA        & -60.144(1)\\
          PBE-GGA        & PZ-LDA           & -60.151(1)\\
          \hline
\end{tabular}
\end{center}
\label{table:bulkDFTNodalQuality}
\end{table}%
The total energy is integrated over the irreducible Brillouin zone by the so-called ``twist-averaging'' procedure:  a DMC calculation is carried out for each symmetry unique $k$-point in a uniform 8x8x8 Monkhorst-Pack mesh and the resulting energies are weight-averaged using by the geometric multiplicity of the k-point as the weight~\cite{Lin:2001fk}.
To treat the finite size errors that occur, both due to the Ewald sums and the finite number of twists, we collect statistics on the static structure factor $S(k)$ during the QMC simulations.  These data are plotted in Figure~\ref{fig:structureFactor}.  The correction to the finite-size errors in the simulation cell's energy is calculated using the functional form for $S(k)$ as detailed by Chiesa \emph{et al.}~\cite{Chiesa:2006uq}.
The 8-atom (triangle symbols) and 16-atom (circle symbols) data for variational Monte Carlo (VMC) and DMC mixed estimators are less well converged when compared to the 54-atom cell.
The corrected DMC mixed estimator (blue) is, however, consistent for all sizes of cell.
This suggests that, at least for some systems, it is possible to estimate the static structure factor correction accurately with data from smaller simulation cells. Since the systems is a simple metal,
within the Random Phase Approximation the behavior of $S(k)$ for small values of $k$ is expected to be proportional to $k^2$, as detail by Bohm and Pines~\cite{Bohm:1953ys}.
The curve fit to our calculated values indicates that $S(k)$ in our simulation is $\propto k^{1.99}$ so that we have reasonable confidence in the quality of the corrected mixed DMC estimator result~\cite{Foulkes:2001uq}.
We use a linear fit to the equation
\begin{equation}
E_n = E_\infty - \frac{a}{N}
\end{equation}
to extrapolate the total energy per atom to the infinite bulk of the cI2 lithium crystal, as shown in Figure~\ref{fig:liCrystalSizeExtrap}.
The data in Figure~\ref{fig:liCrystalSizeExtrap} show that after twist averaging and corrections, the 16 atom cell energy per atom is $\approx~2.5$~mHa from the infinite bulk value, and the 24-atom cell, $\approx~1$~mHa, while the 54 atom is less than a mHa from the extrapolated value. We believe that the Li bulk energy could be further improved by more sophisticated orbitals and optimization by employing more accurate wave functions, such as pfaffians~\cite{Bajdich:2006uq, Bajdich:2008fk}; however, in this study the focus was to understand the trial functions that are, at present, commonly used for solid state and quantum chemistry calculations.

\subsection{Summary}
In order to compare the quality of results for these systems, we estimate the exact energy per atom as the energy of the single atom and the zero point motion (ZPM, infinite nuclear mass) corrected binding energy per atom, in a manner similar to the
previous studies, see for example, Ref.~\cite{Filippi:1996fk}.
For the Li$_4$ cluster, we substitute our own FN-DMC calculated binding energy as the best currently available estimation. 
\begin{table}
\caption{Summary of the estimated total energies per atom for a sequence of different size Li systems. 
$E_\text{tot}$ for $n=4$ cluster we substitute use the obtained FN-DMC value for the binding energy.}
\begin{center}
\begin{tabular}{l l l l}
 \hline\hline
   size  		&$E_\text{FN-DMC}$   & \emph{Est.} $E_0$ & \emph{Estimated from}\\  
  \hline\hline
1      			&$-7.4780(1)$		&$-7.47806	$&Hylleraas expan.\\
2 			&$-7.4976(1)$\footnote{from reference~\cite{Bressanini:2005bs}} &$-7.4977	$&Exp.+ZPM$\footnote{Atom + ZPM corrected exp. binding from Ref.~\cite{Filippi:1996fk}}$\\
4			&$-7.50538(1)$		&$-7.50541	$&FN-DMC binding\\
\hline
cI2 crystal		&$-7.5369(6)$		&$-7.54078	$&Exp.+ZPM\footnote{Atom + ZPM corrected exp. binding from Ref.~\cite{Staikov:1997fk, Gschneidner:1964vn}}
\end{tabular}
\end{center}
\label{table:lithiumPerAtom}
\end{table}%
Table \ref{table:lithiumPerAtom} shows the per atom energy evolving toward the bulk value as the size of the lithium system increases.

The imperfect result for the FN-DMC correlation energy in the Li solid is still very accurate, higher only by about 3 mHa  ($\approx$ 3 \%) from the estimated exact value obtained from the experimental cohesive energy. 
The underestimation of the cohesive energy is essentially the same magnitude, approximately $ 0.1 \, e$V.
We consider that remarkably accurate in light of how simple the single-reference Hartree-Fock
wave function nodes are.
Given the fact that the spin-up and spin-down channels are completely decoupled in HF, the complexity of the nodal surface is not fully captured by this trial function; nevertheless the accuracy of the total energy appears to be quite robust so that the impact of these errors is comparably small.
We conjecture that the electronic structure is dominated by the nearly-free electron picture 
that is far from the strongly correlated regime so that single-reference wave functions lead to qualitatively and also quantitatively accurate descriptions of ground state properties of the Li solid.

\section{Conclusion}
As the size of Li systems increases from a single atom to the bulk crystal, it is clear that the complexity of the nodal hypersurface grows.
In the simplest case of the atom, a nearly exact approximation to the node is known.
For related small systems, the nodal error errors are small, and it is possible to recover almost exact nodes with acceptable sizes of expansions in excited determinants. 
What is valuable and somewhat unexpected, is the fact that the accuracy of the FN-DMC calculation with single reference trial functions is high even for the Li solid.
Note that the solid phase is metallic, so that its electronic structure is different from atomic and molecular systems with localized ground states.
The presented calculations show that for Li systems, readily available trial wave functions 
are sufficiently accurate to provide cohesive and binding energies to within an accuracy of $0.05-0.1 \, e$V.
While our understanding of the fixed-node errors is still limited, the results presented here uncover another piece in the mosaic of previously obtained results which indicate that both the 
electronic density and the complexity of bonds, in particular, the bond multiplicities, strongly influence the nodal accuracy.

\section{Acknowledgements}
We would like to acknowledge the DOE grant DE-SC001231 for support.
We also would like to thank Jind\v rich Koloren\v c for insightful discussions in the early stages of these calculations.

\bibliography{manuscript}{} 
\bibliographystyle{aipauth4-1}

\end{document}